\documentclass[prl,twocolumn,showpacs,preprintnumbers,amsmath,amssymb,aps]{revtex4}
\usepackage{color}
\usepackage{bm, graphicx, amsmath}
\usepackage{hyperref}

\newcommand{\two}[2]{\begin{array}{c}\\[-1.5em]\scriptstyle #1\\[-.5em] \scriptstyle #2\end{array}}

\begin{document}
\title{Wave-particle duality relations based on entropic bounds for which-way information}
\author{Emilio Bagan$^{1}$, J\'anos A. Bergou$^{2,3}$, and Mark Hillery$^{2,3}$}
\affiliation{$^{1}$F\'{i}sica Te\`{o}rica: Informaci\'{o} i Fen\`{o}mens Qu\`antics, Universitat Aut\`{o}noma de Barcelona, 08193 Bellaterra (Barcelona), Spain \\ $^{2}$Department of Physics, Hunter College of the City University of New York, 695 Park Avenue, New York, NY 10065 USA \\ $^{3}$Physics Program, Graduate Center of the City University of New York, 365 Fifth Avenue, New York, NY 10016}
\begin{abstract}

We present wave-particle duality relations involving the relative entropy coherence measure, which plays a prominent role in the resource theory of coherence. The main input in these relations is an entropic bound for the which-way information, which we derive in this letter. We show that this latter crucially depends on the choice of the measurement strategy to obtain the path information. In particular, we present results for two strategies: zero-error identification of the path-detector states, which never produces an error but sometimes fails to return a conclusive answer, and a mixed strategy where both errors and failure are allowed.

\end{abstract}

\maketitle

Coherence is an important resource for quantum information processing. It can be quantified consistently via recently introduced resource theoretical coherence measures. In a seminal work, Baumgratz \emph{et al.,} introduced two such measures,  one involving the entropic distance between a given state and the nearest incoherent state and one involving the trace distance between the same quantities, called the $l_{1}$ measure \cite{baumgratz}. Quantitative wave-particle duality relations, in which the entropic coherence measure is one of the ingredients, are important because of the prominent role of this measure in coherence theory~\cite{Streltsov} and its operational meanings: the relative entropy measure of coherence is identical to the distillable coherence and, for pure states, it coincides with the coherence cost~\cite{Gour,Winter}. In a multipath interferometric scenario, the other ingredient is a quantitative measure of the path distinguishability, which depends on how the path information is obtained. In this paper our main concern is the error-free (also termed unambiguous) identification of the path states, in which no mistakes are made, but the measurement has a probability of failing. We also consider mixed strategies, in which there are both an error and a failure probability. As an important intermediate step towards our main goal, we derive entropic upper bounds for the success probability of both of these strategies. Then, using these entropic bounds, we derive wave-particle duality relations that involve the relative entropy measure of coherence as the main result of this letter.   

Before addressing the details of the derivation, we want to state our central result, the entropic duality relations in an $N$-path interferometer for the case when the path information is obtained via the measurement that results in error-free identification of the path states or, more precisely, the states of the path detector. The success probability of unambiguous discrimination of states  is customarily denoted by $P_{s}$. %{\color{red} One of the results of the paper is that we derive an asymptotically (in~$N$) tight upper bound for the success probability of discriminating $N$ states and show that, surprisingly, it is also the \emph{entropic} measure of distinguishability of the states, i.e., the mutual information between detector outcomes  and path states.} Therefore, in 
In the context of duality we introduce the notation $P_{s}\equiv D$, in order to emphasize that it actually is the distinguishability. The other entry is the entropic measure of coherence. The relative entropy of coherence for a density matrix $\rho$ is given by $C_{\rm{rel-ent}}(\rho)=S(\rho_{\rm{diag}}) - S(\rho)$, where $\rho_{\rm{diag}}$ is a diagonal density matrix in the specified basis whose diagonal elements are the same as those of $\rho$, and $S$ denotes the von Neumann entropy, with the logarithms taken base 2. In our case the relevant density matrix is given later, by Eq. \eqref{rhop}. The normalized entropic coherence measure is $C_{\rm{rel-ent}}(\rho)/\log N$ which we denote by $C$, the coherence. In terms of the distinguishability $D$ and coherence $C$ our entropic duality relation is given as
\begin{equation}
D + C \leq 1 .
\label{duality1}
\end{equation}
Despite its superficial simplicity, this is quite an elaborate relation. It involves finding upper bounds for the relative entropy, carrying the coherence information, and for the mutual information, related to the which-way information. In both cases we make use of Holevo's theorem for the mutual information to establish tight bounds for the involved quantities.

Now we turn our attention to state discrimination which is the underlying problem of obtaining which way information. When discriminating among a set of quantum states, there are several strategies one can adopt.  If the states are not orthogonal, a penalty must be paid, and the choice of strategy is really about the type of penalty to be incurred.  In one case, the minimum-error strategy, one will make mistakes, misidentifying a state, but one finds the procedure that minimizes the probability of this happening \cite{helstrom}. In a second strategy, unambiguous discrimination, one will never make a mistake, but the measurement will sometimes fail, and the object is to find a protocol that minimizes the probability of failure \cite{ivanovic,dieks,peres}.  It is also possible to use a strategy that interpolates between the two, having both an error and a failure probability \cite{chefles,zhang1,fiurasek,eldar,steinberg,hayashi,sugimoto,bergou0}.  For reviews on state discrimination see \cite{bergou1,barnett, bergou2}.

While optimal solutions are known for two states, the situation becomes complicated quickly once the number of states is larger.  There are solutions in special cases, for example, when the states are related by a symmetry group.  For the general case, it is possible to obtain bounds.  We will initially concern ourselves here with unambiguous discrimination, and in that case there are lower bounds on the failure probability \cite{zhang,bandyopadhyay,sun}.  One of these bounds proved useful recently in deriving a wave-particle duality relation \cite{bera}.  Here we would like to derive an entropic lower bound for the failure probability of the unambiguous discrimination of $N$ states, and we will also make use of it to derive a wave-particle duality relation.  We will then proceed to the case in which there is both an error and a failure probability and derive a bound that incorporates  both of these probabilities. Finally, mainly in the supplemental material \cite{supp}, by making use of examples and using numerical analysis based on semidefinite programing, we will present cases for how well the bounds perform.  

We will begin by exploiting the Holevo bound to get a lower bound for the failure probability for unambiguous state discrimination \cite{nielsen}.  In order to do so, we make use of the usual communication scenario in which Alice sends one of the states $\{ |\psi_{x}\rangle \, |\, x=1,2,\ldots, N\}$ to Bob, and the state $|\psi_{x}\rangle$ is sent with probability $p_{x}$.  Bob then applies an unambiguous state discrimination measurement to see which state Alice sent.  Let $X$ be the random variable corresponding to the state Alice sent, and $Y$ be the random variable corresponding to Bob's measurement result.  $X$ takes values in the set $\{ 1,2, \ldots, N\}$ and $Y$ takes values in the set $\{ 1,2, \ldots, N, f\}$, where $f$ corresponds to the measurement failing.  The mutual information between Alice and Bob is 
\begin{equation}
\label{mutual-inf}
I\equiv I(X:Y) = \sum_{x,y}p(x,y)\log \left[ \frac{p(y|x)}{p_{Y}(y)}\right] ,
\end{equation}
where the logarithms are base 2.  
The conditional probabilities, $p(y|x)$, for $X$ and $Y$ are $p(y|x)=\delta_{x,y} p_{xx}$ for $y\neq f$, and $p(f|x)=1-p_{xx}$, where $p_{xx}$ is the probability that if $|\psi_{x}\rangle$ is sent, then $|\psi_{x}\rangle$ is detected.  For the joint distribution, we then have that $p(x,y)= \delta_{x,y} p_{xx} p_{x}$ for $y\neq f$, and $p(x,f)=(1-p_{xx})p_{x}$.  The distribution for $Y$ is given by
\begin{equation}
p_{Y}(y)=\sum_{x}p(x,y) = \left\{ \begin{array}{cc} p_{yy}p_{y}, & y\neq f \\[.5em] \sum_{x} (1-p_{xx})p_{x}, & y=f \end{array} \right. .
\end{equation}
Note that $p_{Y}(f)$ is the total failure probability for the measurement.

We can now compute the mutual information.  We begin with
\begin{equation}
I\! =\!\! \! \sum_{x,y\neq f}\!\! p(x,y)\!\log\! \left[ \frac{p(y|x)}{p_{Y}(y)}\right] 
\!+\! \sum_{x}p(x,f)\!\log\! \left[ \frac{p(f|x)}{p_{Y}(f)}\right] .
\end{equation}
This gives us that
\begin{equation}
I\! =\! -\sum_{x} p_{x}p_{xx}\log p_{x} +\! \sum_{x} p_{x}(1\!-\!p_{xx})\! \log\! \left[ \frac{1\!-\! p_{xx}}{p_{Y}(f)}\right] .
\end{equation}
Now note that
\begin{equation}
\log \left[ \frac{1-p_{xx}}{p_{Y}(f)}\right] = \log \left[ \frac{p_{x}(1-p_{xx})}{p_{Y}(f)}\right] -\log p_{x},
\end{equation}
and that
\begin{equation}
-p_{x}p_{xx}\log p_{x} - p_{x}(1-p_{xx}) \log p_{x} = -p_{x} \log p_{x} .
\end{equation}
Making use of these equations and setting $q_{x}=p_{x}(1-p_{xx})/p_{Y}(f)$, we have that
\begin{equation}
I=H(\{p_{x}\}) + p_{Y}(f) \sum_{x} q_{x}\log q_{x} ,
\end{equation}
where $H(\{p_{x}\})$ is the Shannon entropy of the distribution $\{ p_{x}\}$.  Note that $\sum_{x}q_{x} =1$, and $q_{x}\geq 0$, so that we can express the mutual information as
\begin{equation}
I= H(\{p_{x}\}) - p_{Y}(f)H(\{q_{x}\}) .
\end{equation}

We can now apply Holevo's theorem.  Defining the density matrix, 
\begin{equation}
\rho=\sum_{x}p_{x} |\psi_{x}\rangle\langle\psi_{x}| ,
\end{equation}
the theorem implies that $I \leq S(\rho )$, since Alice is sending pure states, or
\begin{equation}
H(\{q_{x}\}) p_{Y}(f) \geq H(\{p_{x}\}) -S(\rho) ,
\end{equation}
and making use of the fact that $H(\{q_{x}\}) \leq \log N$, we finally have the bound,
\begin{equation}
\label{noerr}
\frac{H(\{p_{x}\}) -S(\rho)}{\log N} \leq  p_{Y}(f) =P_{f} ,
\end{equation}
where, as we noted previously, we took into account that $p_{Y}(f)$ is just the total failure probability for the measurement, $P_{f}$, so this inequality gives us our desired lower bound.  Note that in the case that all of the states are orthogonal, the left-hand side is equal to zero.  This is as expected, of course, since orthogonal states are perfectly distinguishable, and the failure probability is zero.

Next, we want to apply this inequality to wave-particle duality.  We consider a quantum particle, which can travel via $N$ paths through an interferometer.  Each path corresponds to a state $|j\rangle_{p}$, and these states are orthonormal.  There are detectors that measure which path the  particle is in, and if the particle is in path $|j\rangle_{p}$, the detectors are in the state $|\eta_{j}\rangle_{d}$.  The detector states are not, in general, orthogonal.  If they are, we have perfect path information, but if not, then we have some information about the path.  The extent to which we can distinguish the states $|\eta_{j}\rangle_{d}$ forms the basis for quantifying the amount of path information we have.  This approach was pioneered by Englert \cite{englert}, who treated the case of two paths, and used minimum-error state discrimination to distinguish the detector states. Soon after, this was extended in several directions. First, a connection to entanglement was pointed out and concurrence became a part of the duality relation, which then more properly was termed complementarity relation \cite{englert2,jakob1}. Then, it was also extended to include $N$ paths \cite{jakob2}. In all these works minimum-error strategy was used, while here we will be using the zero-error strategy to distinguish the detector states.

The other quantity that appears in a wave-particle duality relation is a quantity related to the ability of the particle to produce an interference pattern, or, alternatively, the ability to use the state of the particle to determine the phases of phase shifters placed in the different paths.  Englert used the visibility of the interference pattern, an $l_{2}$ measure, while more recent approaches, all dealing with $N$ paths, have made use of recently introduced measures of the coherence of the quantum state of the particle \cite{bera,bagan,bagan2}. As already mentioned in the introduction, these coherence measures are a part of the resource theory of quantum coherence developed in \cite{baumgratz}.  In this theory, one chooses a basis, and the coherence measures are defined with respect to this basis.  In our case the path states are the natural basis to choose.  Density matrices that are diagonal in the special basis are taken to be incoherent, and the amount of coherence in a particular state is determined by how far it is from the set of incoherent states.  Two distances were used in \cite{baumgratz}, the $l_{1}$ norm and the relative entropy, each leading to a coherence measure.  The $l_{1}$ measure was used in \cite{bera,bagan2}, while \cite{bagan} made use of both, but for minimum-error discrimination of the detector states.  Here we will be interested in the relative entropy measure to quantify path information when unambiguous (zero-error) discrimination is used. 

The path-detector state of the particle inside the interferometer is given by
\begin{equation}
|\Psi\rangle = \sum_{j=1}^{N}\sqrt{p_{j}} |j\rangle_{p} |\eta_{j}\rangle_{d} ,
\end{equation}
where $p_{j}$ is the probability that the particle is in the $j^{\rm th}$ path.  The path reduced density matrix is 
\begin{equation}
\rho_{p} = {\rm Tr}_{d}(|\Psi\rangle\langle \Psi |) = \sum_{j,k=1}^{N}\sqrt{p_{j}p_{k}}\langle \eta_{k}|\eta_{j}\rangle |j\rangle_{p}\langle k| ,
\end{equation}
and the detector reduced density matrix is
\begin{equation}
\rho_{d}= \sum_{j=1}^{N} p_{j} |\eta_{j}\rangle_{d}\langle\eta_{j}| .
\label{rhod}
\end{equation} 
Because both density matrices come from the same pure state, we have that $S(\rho_{d})=S(\rho_{p})$.  The relative entropy measure of coherence for the path density matrix is given~by
\begin{equation}
C_{\rm{rel-ent}}(\rho_{p})= S(\rho_{p}^{\rm diag}) - S(\rho_{p}) ,
\label{rhop}
\end{equation}
where $\rho_{p}^{\rm diag}=\sum_{j=1}^{N} |j\rangle_{p}\langle j|\rho_{p}|j\rangle_{p}\langle j|$ is the diagonal part, in the path basis, of $\rho_{p}$.  In this case,
$S(\rho_{p}^{\rm diag}) = H(\{ p_{j}\})$,  yielding
\begin{equation}
C_{\rm{rel-ent}}(\rho_{p})= H(\{ p_{j}\}) - S(\rho_{p}) = H(\{ p_{j}\}) - S(\rho_{d}) .
\end{equation}

We now introduce the normalized version of this quantity,
\begin{equation}
C \equiv \frac{C_{\rm{rel-ent}}(\rho_{p})}{\log N} = \frac{H(\{ p_{j}\}) - S(\rho_{d})}{\log N} ,
\label{coherence}
\end{equation}
such that $0 \leq C \leq 1$, so $C$, the coherence, is properly normalized. Furthermore, we can recognize in it the left-hand-side of Eq. \eqref{noerr}. If we further  introduce the success probability of state discrimination as $P_{s} = 1 - P_{f}$, it is clear that $P_{s}$ quantifies the available path information, the so-called path distinguishability. So, we further denote $P_{s}$ as $D$, the distinguishability. 
 
If we now make an unambiguous discrimination measurement of the detector states, the bound in Eq. \eqref{noerr} tells us that, in terms of these quantities, the distinguishability, $D$, must satisfy
\begin{equation}
1 - D \geq \frac{H(\{p_{j}\}) -S(\rho_{d})}{\log N} = \frac{C_{\rm{rel-ent}}(\rho_{p})}{\log N} =C.
\label{bound 1}
\end{equation}
With a slight rearrangement, this is our wave-particle duality relation that was already highlighted in Eq.~\eqref{duality1}.  The distinguishability $D$ characterizes the available which-path information; the larger it is, the more path information we have.  The coherence $C$ tells us how much coherence is present in the path state.  The above relation points to the fine interplay between these quantities; as one increases, the other must decrease, and vice versa.  We also note that a formally identical relation was derived in \cite{bera}, however, with a very different meaning. There, $C$ is the $l_{1}$ measure of coherence while $D$ is not the result of maximizing the mutual information, rather it is the result of finding an upper bound for the success probability of the unambiguous discrimination of $N$ states.  

In order to get a feel for how good the bound is, we will compute it for symmetric detector states.  These states satisfy $\langle \eta_{j}|\eta_{k}\rangle = c$ for $j\neq k$.  If these states are equally probable, that is $p_{j}=1/N$, then the success probability, hence the distinguishability is given by $D = 1 - c$, and one can obtain the coherence, $C$, as a function of the distinguishability, $D$, analytically (see the supplemental material \cite{supp}). Figure \ref{f-1} displays $C$ vs. $D$ for these states, parametrized by the number of paths $N$.

\begin{figure}[htbp]
\begin{center}
\includegraphics[scale=.4]{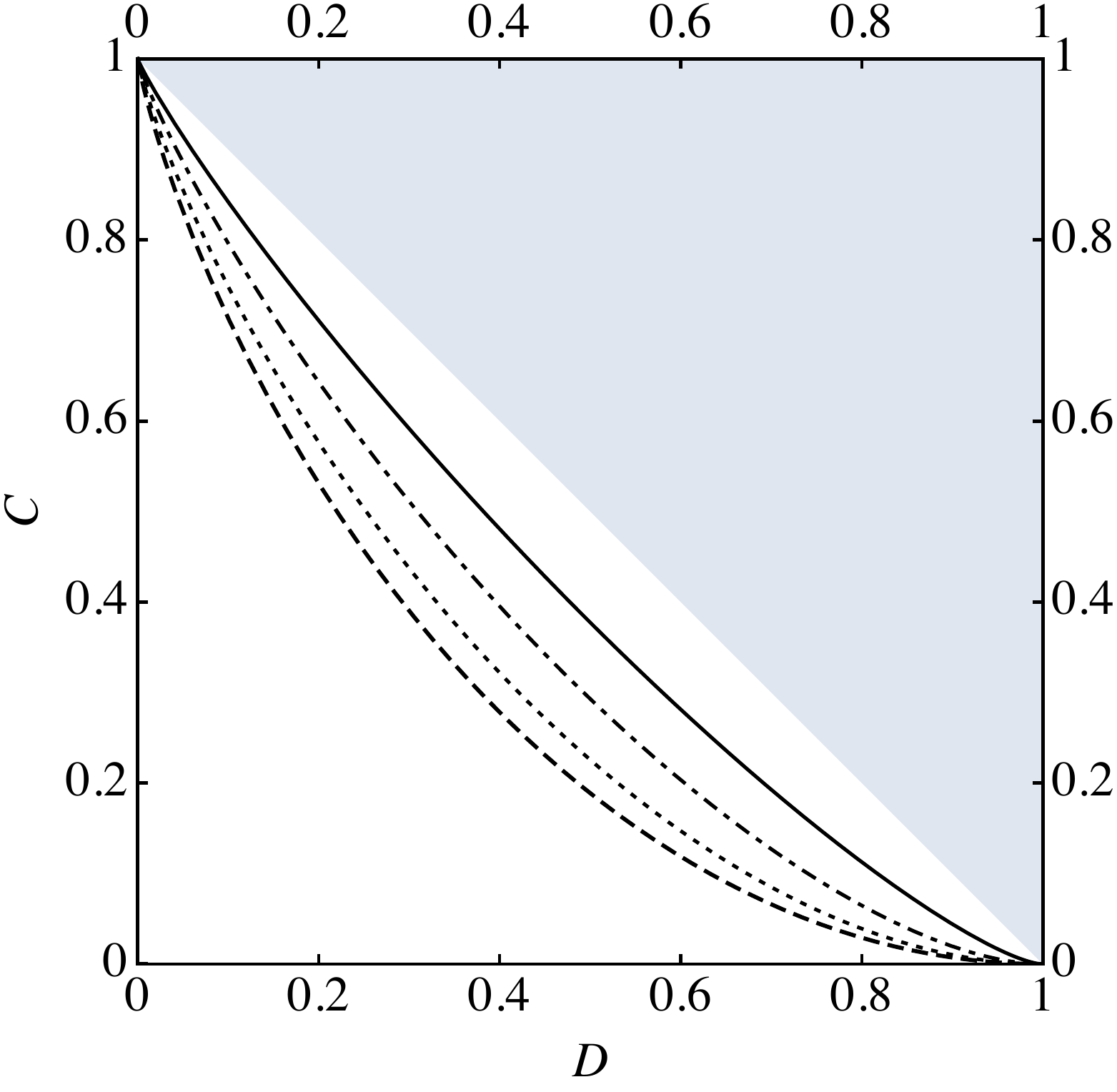}
\caption{Plots of the coherence, $C$, vs. the distinguishability, $D$, for symmetric states of the detector. From bottom to top: dashed line $N=2$; dotted line $N=4$; dot-dashed line $N=16$; solid line $N=256$. The shaded area is the region forbidden by duality, Eq.~(\ref{duality1}).}
\label{f-1}
\end{center}
\end{figure}

Clearly, the bound in Eq.~(\ref{bound 1}) becomes tighter as $N$ increases. With symmetric states, the convergence is only logarithmic in $N$ but, in the Supplemental Material \cite{supp}, we show that  with other states it can be improved to~$1/N$.

We now want to extend our results to include the possibility of making a mistake, so we will no longer assume that $p(y|x)=0$ for $x\neq y$.  We begin by splitting the sum in Eq.\ (\ref{mutual-inf}) into two parts, $y\neq f$ and $y=f$.  Let
\begin{eqnarray}
S_{1} & = & \sum_{x,y \neq f} p(x,y) \log \left[ \frac{ p(y|x)} {p_{Y}(y)} \right], \nonumber \\
S_{2} & = & \sum_{x} p(x,f) \log \left[ \frac{ p(f|x)}{p_{Y} (f)} \right] .
\end{eqnarray}
We can then express these sums as 
\begin{eqnarray}
S_{1} & = & \sum_{x,y\neq f} p(y|x)p_{x} \log \left[ \frac{p(y|x)p_{x}}{p_{Y}(y)}\right] \nonumber \\
 & & -\sum_{x} [1-p(f|x)] p_{x} \log p_{x} , \nonumber \\
S_{2} & = & \sum_{x} p(f|x) p_{x} \log \left[ \frac{p(f|x) p_{x}}{p_{Y}(f)}\right]  \nonumber \\
 & & - \sum_{x} p(f|x) p_{x}\log p_{x} .
\end{eqnarray}

Let us deal with $S_{2}$ first.  The second term will cancel when we add $S_{1}$ and $S_{2}$, so we need only consider the first term.  Define a random variable, $Z$, taking values in the set $\{ 1,2, \ldots N\}$, such that
\begin{equation}
p(Z=z)= \frac{p_{z}p(f|z)}{p_{Y}(f)} .
\end{equation}
Then the first term of $S_{2}$ is just $-H(Z) p_{Y}(f)$, and we note that $p_{Y}(f)$ is just the failure probability for the measurement, which we shall denote as $P_{f}$.  We also note that $H(Z)\leq \log N$.

In order to estimate $S_{1}$ we will make use of Fano's theorem \cite{nielsen}.  This theorem was used to find a lower bound for the error probability in the minimum-error state discrimination of two states in \cite{nielsen}. In order to use the theorem, we will define two new random variables, $\tilde{X}$ and $\tilde{Y}$ both of which take values in the set $\{ 1,2,\ldots N\}$.  The probability that $\tilde{Y} = y$, which we denote as $\tilde{p}(y)$, is given by
\begin{equation}
\tilde{p}(y) = \frac{p_{Y}(y)}{1-P_{f}} .
\end{equation}
For $\tilde{X}$, we define the conditional probability that $\tilde{X}=x$ given that $\tilde{Y}=y$, $\tilde{p}(x|y)$, by
\begin{equation}
\tilde{p}(x|y) = \frac{p(y|x)p_{x}}{p_{Y}(y)} .
\end{equation}
The first term in $S_{1}$ can then be expressed as
\begin{multline}
\sum_{x,y\neq f} (1-P_{f}) \tilde{p}(y) \tilde{p}(x|y) \log \tilde{p}(x|y) \\ = -(1-P_{f}) H(\tilde{X}|\tilde{Y}) .
\end{multline}
If $\tilde{P}_{e}$ is the probability that $\tilde{X}\neq \tilde{Y}$, then Fano's theorem gives us that
\begin{equation}
H_{2}(\tilde{P}_{e}) + \tilde{P}_{e} \log (N-1) \geq H(\tilde{X}|\tilde{Y}) ,
\end{equation}
where $H_{2}(q)= -q\log_{2}q - (1-q)\log_{2}(1-q)$.  Adding $S_{1}$ and $S_{2}$, and noting that according to Holevo's theorem, the result must be less than or equal to $S(\rho )$, we have that 
\begin{eqnarray}
(1-P_{f}) [H_{2}(\tilde{P}_{e}) &+& \tilde{P}_{e} \log (N-1)] + P_{f} \log N   \nonumber \\
&\geq &H(\{p_{x}\}) - S(\rho ) .
\end{eqnarray}
We next need to relate $\tilde{P}_{e}$ to the probability that $X\neq Y$, which we shall call $P_{e}$, and is the error probability of the measurement.  Now
\begin{eqnarray}
\tilde{P}_{e} & = & \sum_{x, y\neq f,x} \tilde{p}(x|y)\tilde{p}(y) \nonumber \\
 & = & \sum_{x,y\neq f,x} \frac{p(y|x) p_{x}}{p_{Y}(y)} \frac{p_{Y}(y)}{1-P_{f}} =\frac{P_{e}}{1-P_{f}} .
\end{eqnarray}
where we have used the fact that
\begin{equation}
P_{e}=\sum_{x,y\neq f,x} p(y|x) p_{x} .
\end{equation}
Making use of this relation, we obtain
\begin{eqnarray}
\label{werr} 
(1-P_{f}) H_{2}\left( \frac{P_{e}}{1-P_{f}}\right) &+& P_{e} \log (N-1) + P_{f} \log N \nonumber \\
&\geq& H(\{p_{x}\}) - S(\rho ) .
\end{eqnarray} 
For $P_{e}=0$, this reduces to the inequality in Eq.\ (\ref{noerr}).  Nonzero values of $P_{e}$ allow $P_{f}$ to be smaller.  There is, therefore, a tradeoff between error probability and failure probability.

Eq.\ (\ref{werr}) can be applied to the case of a particle going through an $N$-path interferometer, and the result is a wave-particle duality relation involving both failure and error probabilities for the detectors.  The derivation is straightforward, with the result  
\begin{eqnarray}
\frac{1\!-\!P_{f}}{ \log N} H_{2}\!\left( \frac{P_{e}}{1\!-\!P_{f}}\right)\! &+&\! P_{e}\frac{\log (N-1)}{\log N} + P_{f} \nonumber \\
&\geq& \frac{C_{\rm{rel-ent}}(\rho_{p})}{\log N} .
\label{frioduality}
\end{eqnarray}
The path information is now characterized by the two probabilities, $P_{f}$ and $P_{e}$, which refer to measurements made on the detector states.  Perfect path information would imply that both are equal to zero, and thus the coherence vanishes, $C_{\rm{rel-ent}}(\rho_{p})=0$. At the other extreme, if the coherence is maximal, $C_{\rm{rel-ent}}(\rho_p)/\log N=1$, one has $P_e/(1-P_f)=1-1/N$ and $P_s/(1-P_f)=1/N$, where $P_s=1-P_e-P_f$ is the success probability. These ratios are the error and the success probability conditioned on no failure, respectively. The values, $1-1/N$ and $1/N$, correspond to random guessing the path. So, there is no path information if the coherence is maximal. We also note that we can get a somewhat simpler bound by noting that $\log (N-1) < \log N$, which allows simplifying Eq.~\eqref{frioduality} to
\begin{equation}
C + D \leq 1+ \frac{1-P_{f}}{\log N} H_{2}\left( \frac{P_{e}}{1-P_{f}}\right)   .
\end{equation}
This still has the right limit as $P_{e} \rightarrow 0$ and differs from Eq. \eqref{duality1} only slightly.  

\emph{Summary} This letter provides the missing links in the study of entropic duality relations. In the case when the path information is obtained from measurements that~have a failure or inconclusive outcome, % with more outputs than the number of paths, $N$, 
such relations have never been studied before. %Measurements to obtain path information with $N$ outputs involve, e.g., the minimum error and the maximum information gain strategies. 
%Here, we extended the study of entropic duality bounds to two more cases. 
Two possible scenarios have been considered. In the first, unambiguous (zero-error) path information is obtained. The measurement never returns an error but sometimes returns an inconclusive answer. In the second, both errors and inconclusive outcomes are permitted. %This measurement interpolates between the unambiguous and the minimum error strategies. In both of these strategies, unambiguous or interpolation, the which way detector has more outcomes than the number of paths, in order to accommodate the inconclusive results. 
We derived entropic lower bounds for these two discrimination problems, and employed them to obtain our main results, the duality relations involving the relative entropy measure of coherence. 
%The first one is for the lower bound on the failure probability for unambiguous discrimination. The second is for a mixed strategy, with both a failure and an error probability, and the bound involves both. 
%Thus, our work completes the study of entropic duality relations and the results can be applied to assessing the information content of quantum states and to quantum information processing, in general.

\emph{Acknowledgments} EB acknowledges support from the Spanish MINECO, project FIS2016-80681-P, and from the Catalan Government, projects CIRIT 2017-SGR-1127 and QuantumCAT 001-P- 001644 (RIS3CAT comunitats), co-financed by the European Regional Development Fund (FEDER).

\section{Supplementary material for Entropic bounds for state discrimination with applications to wave-particle duality}

In order to assess the tightness of the bounds/duality relations in the body of the letter, in particular, 
\begin{equation}
P_{f} \geq \frac{H(\{p_{j}\}) -S(\rho_{d})}{\log N} = \frac{C_{\rm rel.-ent.}(\rho_{p})}{\log N} 
\label{bound 1}
\end{equation}
[Eqs.~(1) and~(12) in the letter] and
\begin{multline}
{1\!-\!P_{f}\over \log N} H_{2}\!\left( \frac{P_{e}}{1\!-\!P_{f}}\right)\! +\! P_{e} {\log (N-1)\over\log N} + P_{f} \\
\geq {C_{\rm rel.-ent.}(\rho_{p})\over\log N}
\label{dual err marg}
\end{multline}
[Eqs.~(31) and~(32) in the letter], we will examine several cases, Secs.~\ref{sec_1} and~\ref{sec_2}, in which we can find the failure and error probabilities either exactly or numerically. In Sec.~\ref{sec-SDP}, we present in some detail the SDP formulation used to obtained our numerical results. 

\subsection{Symmetric states}\label{sec_1}

We will first consider a highly symmetric situation where the probabilities of finding the particle in each path of the interferometer are equal, $p_j=1/N$, and the detector states have equal overlap with each other, namely, $\langle\eta_j|\eta_k\rangle=c$, for $j\not= k$. Then,
\begin{eqnarray}
\rho_p\!\!\!&=&\!\!\!{1\over N}\openone\!+\!{c\over N}\!\!\!\sum_{\two{j,k=1}{j\not=k}}^N\!\! |j\rangle\langle k|
\!=\!
{1\!-\!c\over N}\openone\!+\!{c\over N}\!\!\sum_{j,k=1}^N\!\!|j\rangle\langle k|\nonumber\\
\!\!\!&=&\!\!\!
{1\!-c\!\over N}\openone\!+\!c|f_0\rangle\langle f_0|,
\label{rho_p Fourier}
\end{eqnarray}
where $|f_0\rangle$ is the first element of the Fourier basis, $\{|f_r\rangle\}_{r=0}^{N-1}$, defined as
\begin{equation}
|f_r\rangle={1\over\sqrt N}\sum_{j=1}^N {\rm e}^{i{2\pi\over N}rj}|j\rangle.
\end{equation}
It is then apparent that $\rho_p$ is diagonal in the Fourier basis. The eigenvalue $\lambda_0$, corresponding to the eigenvector $|f_0\rangle$, can be red off from Eq.~(\ref{rho_p Fourier}), and is $\lambda_0=[1+(N-1)c]/N$. Likewise, for $r=1,2,\dots, N$, we have
%%
%\begin{eqnarray}
%\rho_p|f_r\rangle&=&{1\over\sqrt N}\sum_{j=1}^n\left({{\rm e}^{i{2\pi\over N}rj}\over N}+{c\over N}\sum_{\two{k=1}{k\not=j}}^N{\rm e}^{i{2\pi\over N}rk}\right)|j\rangle\nonumber\\
%&=& 
%{1\over\sqrt N}\sum_{j=1}^n\left({{\rm e}^{i{2\pi\over N}rj}\over N}-{c\over N}{\rm e}^{i{2\pi\over N}rj}\right)|j\rangle
%\nonumber\\
%&=&{1-c\over N}|f_r\rangle,
%\end{eqnarray}
%%
%where we have used that the $N$th-roots of unity add up to zero. This relation shows that 
$\lambda_r=(1-c)/N$, for $r=1,2,\dots, N$. Recalling that $S(\rho_p)=H(\{\lambda_r\})$, it is straightforward to compute the entropic coherence measure of $\rho_p$, which can be written as 
\begin{align}
C_{\rm rel.-ent.}(\rho_p)&={(N-1)(1-c)\over N}\log(1-c)\nonumber\\
&+{1+(N-1)c\over N}\log\left[1+(N-1)c\right].
\label{C_rel-ent symm}
\end{align}

For unambiguous discrimination of paths ($P_e=0$), we next need to compute the failure probability $P_f$. To this end, we compute the Gram matrix, $G$, of the states of the detector, whose entries are $G_{jk}=\langle\eta_j|\eta_k\rangle$. So, \mbox{$G_{jk}=c$}, if $j\not=k$ and $G_{jj}=1$, for $j,k=1,2,\dots, N$. The failure probability is known to be given by $P_f=1-\lambda_{\rm min}$, where~$\lambda_{\rm min}$ is the smallest eigenvalue of the Gram matrix~$G$~\cite{secamu-ta,hoeski}. We note that $G$ is actually the matrix $N \rho_p$. Hence, the eigenvalues are $N\lambda_r$, where $\lambda_r$ are the eigenvalues of $\rho_p$ computed above. Therefore, $\lambda_{\rm min}=1-c$, and we have
\begin{equation}
P_f=c.
\label{P_f symm}
\end{equation}
%
%
%\begin{figure}[htbp]
%\begin{center}
%\includegraphics[scale=.4]{symmetric_states4.pdf}
%\caption{In solid lines, plots of $C_{\rm rel.-ent.}(\rho_p)/\log N$ vs. $P_f$ for symmetric states of the detector. From bottom to top, $N=2,4,16,256$. The white area is the region allowed by the duality relation in Eq.~(\ref{bound 1}).}
%\label{f-1.1}
%\end{center}
%\end{figure}
%
Eqs.~(\ref{C_rel-ent symm}) and~(\ref{P_f symm}), enable us to plot $C_{\rm rel.-ent.}/\log N=C$ as a function of $1-P_f=D$ for this particular type of state of the interferometer (Fig.~1 in the body of the letter). %shows such plots for $N=2,4,16,256$ (from bottom to top) in solid lines. The dashed line is the edge of the region (white area) allowed by duality according to Eq.~(\ref{bound 1}). We note that the bound in Eq.~(\ref{bound 1}) becomes tighter as $N$ increases. Moreover, we 
Note that
\begin{equation}
{C_{\rm rel.-ent.}(\rho_p)\over\log N}\sim P_f+O(1/\log N)
\label{O 1/l}
\end{equation}
as $N\to\infty$,
which means that the type of state under consideration saturates the bound/duality relation asymptotically. This behavior is in contraposition to that of the bound based on the $l_1$ measure of coherence~\cite{bagan1}, for which attainability happened only for $N=2$, whereas the bound became looser  for increasing number of paths. 

\begin{figure}[h]
\begin{center}
\includegraphics[scale=.31]{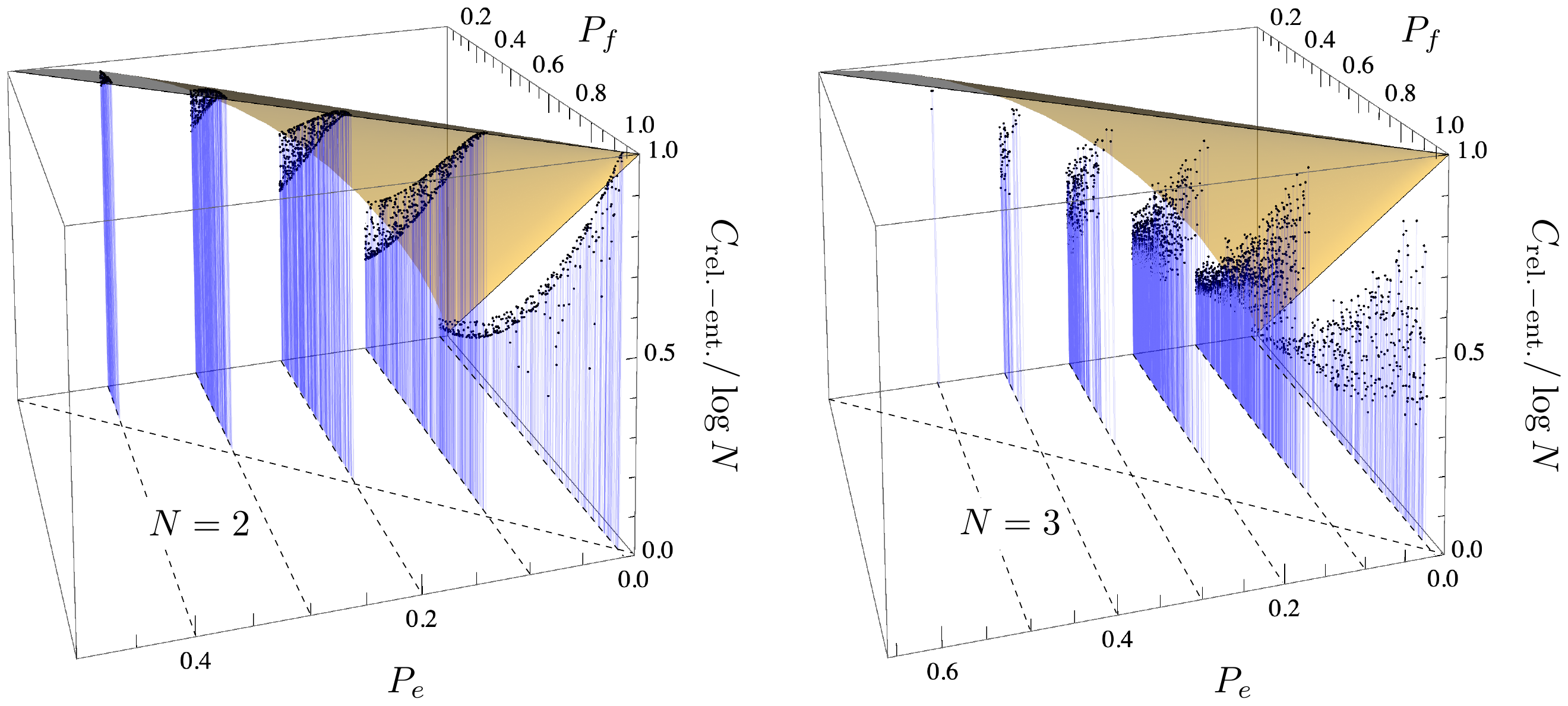}
\caption{On the left (right), the black dots correspond to $C_{\rm rel.-ent.}(\rho_p)/\log N$ as a function of $P_f$ and $P_e$ for random configurations of the 2-path (3-path) interferometer. The brownish translucent surface is given by the left hand side of the duality bound in Eq.~(\ref{dual err marg}). The rightmost bunch of black dots in both figures correspond to $P_e=0.01$, wheres for the others, the value of $P_e$ can be read off from the axis of the plots.}
\label{f-3}
\end{center}
\end{figure}

We next focus on Eq.~(\ref{dual err marg}). To assess its tightness, we have resorted in both numerical and analytical approaches. For $N=2,3$,  we have generated random Gram matrices (or equivalently, $N\rho_p$ states)  and used the highly efficient SDP formulation introduced in Section~\ref{sec-SDP} to compute $P_f$ for fixed maximum allowed error probability $P_e$. The results are displayed in Figure~\ref{f-3}. The left hand side of Eq.~(\ref{dual err marg}) gives the brownish translucent surface and the black dots correspond to the random generated Gram matrices (the vertical blue lines underneath each dot are just to guide the eye). In the left (right) figure, we have chosen from right to left $P_e=0.01,0.1,0.2,0.3,0.4$ (and $0.5$).  The plots show that the larger the value of $P_f$, the tighter the bound/duality relation becomes.

For $N>3$, it is hard to generate random Gram matrices in the region of interest (i.e., near the brownish surface of the figures), so we again considered symmetric states, as we did for unambiguous path discrimination.
If the maximum allowed error probability is in the range $0\le P_e\le P_e^{\rm min}$, where $P_e^{\rm min}$ is the minimum error when no inconclusive/failure outcome is allowed (minimum error discrimination protocol), the maximum success probability is known to be~\cite{ulrike}
\begin{equation}
P_s={1\over N}\!\left[\sqrt{{1\!+\!(N\!-\!1)c\over N}\!-\!P_f}\!+\!(N\!-\!1)\sqrt{1\!-\!c\over N}\,\right]^2.
\label{frio}
\end{equation}
Since $P_s+P_e+P_f=1$, we can substitute $P_s$ by $1-P_e-P_f$ in Eq.~(\ref{frio}) and solve for $P_f$ to obtain
\begin{equation}
P_f=c-2\sqrt{{1-c\over N-1}P_e}+{N\over N-1}P_e,
\end{equation}
where we note that for $P_e=0$ we recover Eq.~(\ref{P_f symm}). Using this expression and Eq.~(\ref{C_rel-ent symm}) it is not hard to check that
\begin{equation}
{C_{\rm rel.-ent.}(\rho_p)\over \log N}\sim P_f+P_e+O(\log N)
\end{equation}
for large $N$. This asymptotic behavior coincides with that of the left hand side of Eq.~(\ref{dual err marg}). Hence, this bound becomes tighter as $N$ increases.

\subsection{Asymmetric states}\label{sec_2}

\begin{figure}[tbp]
\begin{center}
\includegraphics[scale=.35]{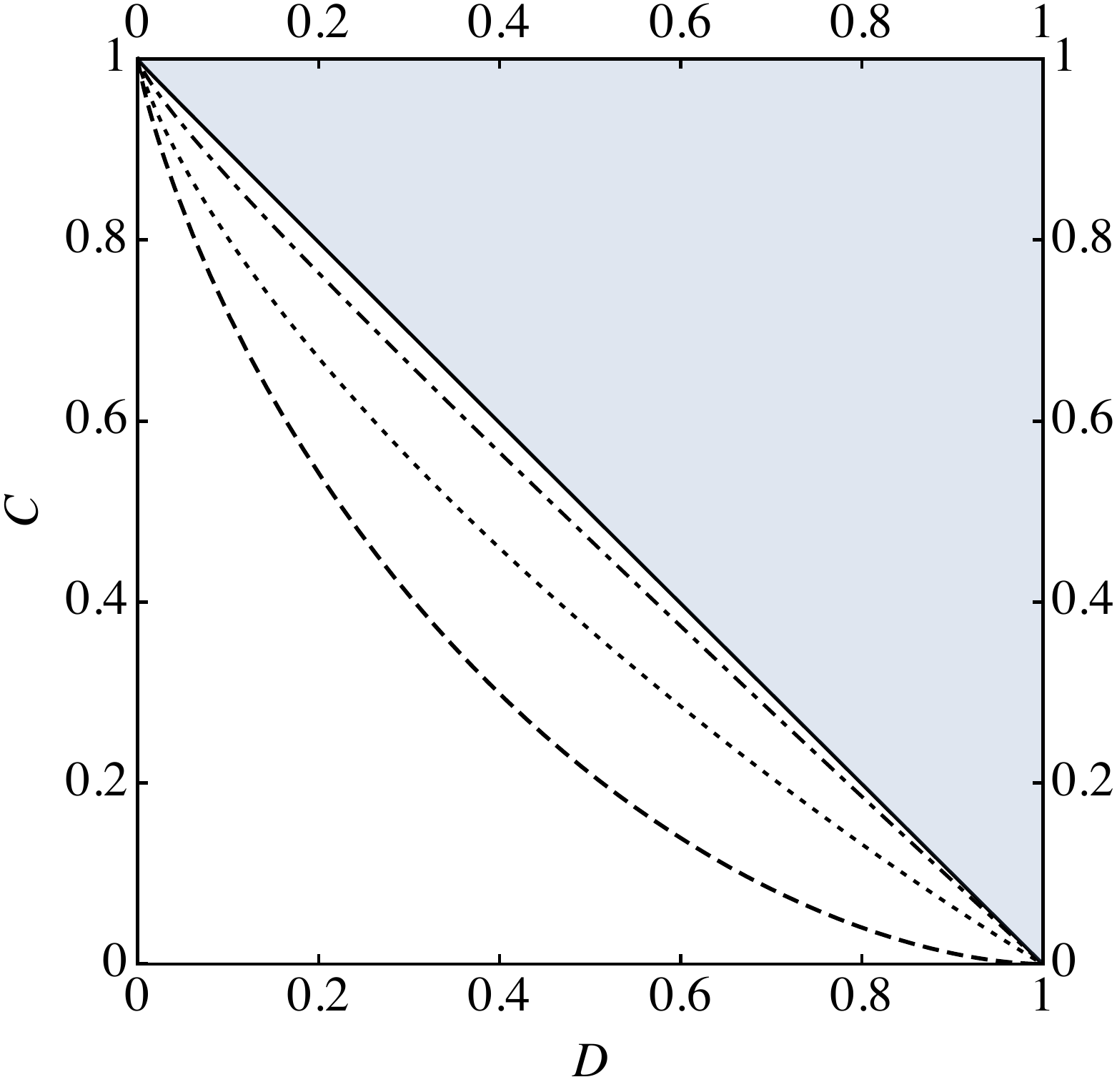}
\caption{Plots of the coherence, $C$, vs. the distinguishability, $D$, for assymmetric states of the detector. From bottom to top: dashed line $N=2$; dotted line $N=4$; dot-dashed line $N=16$; solid line $N=256$. The latter is almost indistinguishable from the boundary of region (shaded area) forbidden by duality, Eq.~(\ref{bound 1}).}
\label{f-2}
\end{center}
\end{figure}

We can improve the previous result in Eq.~(\ref{O 1/l}) by considering a  state $\rho_p$ of the  form %Gram matrix of the form 
%%
%\begin{equation}
%G_{ij}=\left\{
%\begin{array}{lcl}
%p,&\ & \mbox{\rm if $i=j=1$,}\\[.5em]
%\displaystyle {1-p\over n-1}, &&\mbox{otherwise.}
%\end{array}
%\right.
%\end{equation}
%%
%
\begin{equation}
\rho_p=p|N\rangle_p\langle N|+{1-p\over N-1}\sum_{j,k=1}^N\kern-.3em\raisebox{.4em}{$'$}|j\rangle_p\langle k |,\ \ {1\over N}\le p\le 1,
\label{rho_p 2}
\end{equation}
where the prime means that the case $j=k=N$ is excluded from the double sum. This state corresponds to the choice
\begin{equation}
\langle \eta_j|\eta_N\rangle=\sqrt{1-p\over p(N-1)},\quad j=1,2,\dots, N-1,
\end{equation}
and $\langle \eta_j|\eta_k\rangle=1$, for $j\not=k$ and $j$, $k$ not both equal to~$N$. In other words, $|\eta_j\rangle=|\phi\rangle$, $j=1,2,\dots,N-1$,  and thus the paths $1$, $2$, $\dots$, $N-1$ cannot be distinguished by~the states of the detectors. They can only distinguish (to some degree) the $N$th path from the others. The probability of finding the particle in one of these indistinguishable paths is  $q=(1-p)/(N-1)$, whereas the probability of finding the particle in the $N$th path is $p$.  In this situation, the optimal unambiguous (zero-error) discrimination protocol consists in performing the von Neumann measurement $\{ \Pi_N=|\phi^\perp\rangle\langle\phi^\perp|,\Pi_f=|\phi\rangle\langle\phi|\}$, whose failure probability~is
\begin{eqnarray}
P_f&=&\sum_{j=1}^{N-1} p_j|\langle\eta_j|\phi\rangle|^2+p_N|\langle\eta_N|\phi\rangle|^2\nonumber\\
&=&1-p+{1-p\over N-1}= {N(1-p)\over N-1}.
\label{Pf 2}
\end{eqnarray}
To compute the coherence of $\rho_p$, we need to find its eigenvalues. We note that there exists (unnormalized) eigenvectors of $\rho_p$ of the form
\begin{equation}
|u\rangle=a|N\rangle+\sum_{j=1}^{N-1}|j\rangle=a|N\rangle+\sqrt N|\tilde f_0\rangle,
\end{equation}
where $|\tilde f_0\rangle$ is the first element of the Fourier basis $\{|f_r\rangle\}_{r=0}^{N-2}$ for the span of $\{|j\rangle\}_{j=1}^{N-1}$. Substituting $|u\rangle$ into Eq.~(\ref{rho_p 2}) we find
\begin{equation}
\rho_p|u\rangle\!=\!\left(ap\!+\!1\!-\!p\right)|N\rangle\!
+\!\left(\!{1\!-\!p\over N\!-\!1}a\!+\!1\!-\!p\!\right)\sqrt N|\tilde f_0\rangle.
\end{equation}
So, we must have
\begin{eqnarray}
ap+1-p&=&\lambda a, \\
{1-p\over N-1}a+1-p&=&\lambda.
\end{eqnarray}
Solving this system one obtains the two eigenvalues
\begin{equation}
\lambda_{\pm}={1\pm \sqrt{1-{N-1\over N}P_f(1-P_f)}\over2},
\end{equation}
where we have used Eq.~(\ref{Pf 2}).
It is straightforward to check that $|\tilde f_r\rangle$, $r=1,2,\dots, N-2$ are also eigenvectors of $\rho_p$ with zero eigenvalue, therefore
\begin{eqnarray}
C_{\rm rel.-ent.}(\rho_p)&=&-p\log p-(1-p)\log{1-p\over N-1}\nonumber\\
&+&H(\{\lambda_+,\lambda_-\}).%\nonumber\\
%&=&-\left(1-{N-1\over N}P_f\right)
\end{eqnarray}
Using Eq.~(\ref{Pf 2}) again, one can trade $p$ for $1-P_f=D$ and obtain $C_{\rm rel.-ent.}(\rho_p)/\log N=C$ as a function of $D$. Plots of this function are shown in Fig.~\ref{f-2} for $N=2,4,16, 256$ (bottom to top). 
Note that as $N$ increases the lines approach the boundary of the allowed region, given by the bound in Eq.~(\ref{bound 1}), much faster than the corresponding lines in Fig.~1 of the letter. Actually, one can check that
\begin{equation}
C\sim 1-D+O(1/N),
\end{equation}
for these very asymmetric states. This observation shows that the duality bound in Eq.~(\ref{bound 1}) is very tight already for relatively small values of $N$.  However, for $P_e>0$, the subleading correction becomes again $O(1/\log N)$. 

\subsection{Semidefinite program}\label{sec-SDP}

In this section we set up unambiguous discrimination and error margin discrimination as SDP problems. Let $\{|\psi_j\rangle\}_{j=1}^N$ be the set of states, not necessarily independent, that we wish to discriminate. %We will assume that the $N$  pure stated that we wish to discriminate are independent. %If this is so, the optimal unambiguous protocol can be written as~\cite{secamu-ta}
%%
%\begin{equation}
%\min_{\Gamma\ge 0} \left(1-{\rm tr}\ \Gamma\right)\quad\mbox{\rm subject to}\quad
%G-\Gamma_{\rm diag}\ge 0,
%\end{equation}
%%
%where $\Gamma$ is a $n\times n$ positive matrix, $\Gamma_{\rm diag}$ is the diagonal matrix consisting of the diagonal entries of $\Gamma$, and $G$ is 
%The Gram matrix of this set of states $\{|\psi_j\rangle\}_{j=1}^N$. 
Given any orthonormal basis $\{|j\rangle\}_{j=1}^N$, the Gram matrix of this set can be written as
\begin{equation}
G=\sum_{j,k=1}^N \sqrt{p_jp_k}\langle\psi_j|\psi_k\rangle\;|j\rangle\langle k|.
\end{equation}
This matrix encodes all the information we need for the problem at hand.
%If $\Gamma^*$ attains the minimum, then the failure probability is given by
%%
%\begin{equation}
%P_f=1-{\rm tr}\, \Gamma^*.
%\end{equation}
%%

Let us first consider the error margin discrimination problem whereby we tolerate some identification errors. More precisely, we allow an error probability no larger than some prescribed value $P_e$. In particular, if \mbox{$P_e=0$}, the protocol reduces to unambiguous discrimination.
Larger values of $P_e$ will obviously result in a smaller value of the minimum failure probability $P_f$. For~$P_e$ larger than the optimal minimum-error discrimination probability we have $P_f=0$.

The error margin discrimination problem we have just discussed can be formulated as the following SDP optimization. Let ${\mathcal P}$ stand for the set of positive semidefinite block-diagonal matrices 
\begin{eqnarray}
Z:&=&\bigoplus_{j=1}^N z_j,\\
&&\nonumber
\end{eqnarray}
where each block, $z_j$, $j=1,2,\dots, N$, is $N\times N$ (note that $Z\ge0$ iff $z_j\ge0$ for all $j$), and let $B$ be the (block-diagonal) matrix defined by
$$
B=\bigoplus_{j=1}^N b_j,\quad b_j:=\openone-|j\rangle\langle j|
$$
(i.e., $B$ is a diagonal matrix whose entries are zero or one; a zero on each block). Then, we can compute the $Z$ matrix that attains  
\begin{equation}
\min_{Z\in{\mathcal P}} \left(1-{\rm tr} Z\right)\quad\mbox{\rm subject to}\quad
\left\{
\begin{array}{l}
\displaystyle G-\sum_{j=1}^N z_j\ge 0,\\[1.5em]
{\rm tr}\left(ZB\right)\le P_e.
\end{array}
\right.
\label{err marg sdp}
\end{equation}
If $Z^*$ is such matrix, the failure probability is given by
\begin{equation}
P_f=1-{\rm tr}\ Z^*,
\end{equation}
provided that $P_e$ is smaller than the optimal minimum-error probability.
%In Eq.~(\ref{err marg sdp}), $\{E_j\}_{j=1}^N$ is a set of $N\times N$ hermitian matrices,  $Z$ is the block-diagonal matrix
%%
%\begin{eqnarray}
%Z:&=&\bigoplus_{j=1}^N E_j,\\
%&&\nonumber
%\end{eqnarray}
%%
%(note that $Z\ge0$ imply $E_j\ge0$ for $j=1,2,\dots,N$), and the matrix $B$ is defined by
%$$
%B=\bigoplus_{j=1}^N B_j,\quad B_j:=\openone-|j\rangle\langle j|,
%$$
%i.e., $B$ is a diagonal matrix whose entries are zero or one.
Unambiguous discrimination is a particular case of these equations where we set $P_e=0$. Alternative formulations of unambiguous discrimination can be found in~\cite{secamu-ta}.

Eq.~(\ref{err marg sdp}) can be derived as follows. Let $\{\Pi_j\}_{j=1}^N$ be a set of positive operators such that
\begin{equation}
\sum_{j=1}^N \Pi_j\le\openone.
\label{sum pi =1}
\end{equation}
 Then, $\Pi_0:=\openone-\sum_{j=1}^N \Pi_j\ge 0$ can be though of as the failure POVM operator (i.e., the operator that gives the inconclusive outcome), whereas a click of $\Pi_j$ identifies $|\psi_j\rangle$ for $j=1,2,\dots, N$. Let us define the unnormalized states $|\tilde\psi_j\rangle:=\sqrt p_j |\psi_j\rangle$ and let $\Gamma$ be the matrix
\begin{equation}
\Gamma:=\sum_{k=1}^N |\tilde\psi_k\rangle\langle k| .
\end{equation}
%
%where $\{|k\rangle\}_{k=1}^N$is any orthonormal basis of the span of~$\{|\psi_j\rangle\}_{j=1}^N$. 
Multiplying Eq.~(\ref{sum pi =1}) by $\Gamma^\dagger$ on the left and by $\Gamma$ on the right and defining $z_j:=\Gamma^\dagger\Pi_j \Gamma\ge0$, we obtain
\begin{equation}
\sum_{j=1}^N z_j=\sum_{j=1}^N \Gamma^\dagger\Pi_j \Gamma\le \Gamma^\dagger \Gamma=G,
\end{equation}
which is the first condition in Eq.~(\ref{err marg sdp}). We next compute the trace of~$Z$ 
\begin{eqnarray}
{\rm tr}\, Z&=&\sum_{j=1}^N {\rm tr}\ z_j=\sum_{j,k=1}^N\langle\tilde\psi_k|\Pi_j|\tilde\psi_k\rangle\nonumber\\
&=&1-P_f.
\end{eqnarray}
Hence, $P_f=1-{\rm tr}\ Z$. Likewise, %writing $B_j=\openone-|j\rangle\langle j|$, 
we have
\begin{eqnarray}
{\rm tr}\left(ZB\right)&=&\sum_{j=1}^N\left({\rm tr}\ z_j-\langle j|z_j|j\rangle\right)\nonumber\\
&=&\!\!\!\sum_{\two{j,k=1}{k\not=j}}^N\langle\tilde\psi_k|\Pi_j|\tilde\psi_k\rangle
\le P_e.
\end{eqnarray}
This is the second condition in Eq.~(\ref{err marg sdp}) and concludes the derivation.

\end{document}